# On Tangible User Interfaces, Humans and Spatiality

Ehud Sharlin[1], Benjamin Watson[2], Yoshifumi Kitamura[1], Fumio Kishino[1], Yuichi Itoh[1]

*[1] Human Interface Engineering Laboratory, Osaka University, Japan; [2] Department of Computer Science, Northwestern University, USA*

Abstract: Like the prehistoric twig and stone, tangible user interfaces (TUIs) are objects manipulated by humans. TUI success will depend on how well they exploit spatiality, the intuitive spatial skills humans have with the objects they use. In this paper we carefully examine the relationship between humans and physical objects, and related previous research. From this examination we distill a set of observations, and turn these into heuristics for incorporation of spatiality into TUI application design, a cornerstone for their success. Following this line of thought, we identify "spatial TUIs", the subset of TUIs that mediate interaction with shape, space and structure. We then examine several existing spatial TUIs using our heuristics.

*Keywords: tangible user interfaces, design heuristics, spatial mappings, affordances.*

## 1. Introduction

Like other HCI technologies, tangible user interfaces (TUIs) strive to increase human productivity by making their digital tools easier to use. TUIs achieve this by exploiting human spatiality, our innate ability to act in physical space and interact with physical objects. The desktop mouse is a powerful and early example of the impact this approach can have on HCI and productivity.

Fitzmaurice et al. were the first to distinguish TUIs from other interfaces – though they called them "graspable" user interfaces [4]. Fitzmaurice defined a graspable user interface as: "a physical handle to a virtual function where the physical handle serves as a dedicated functional manipulator" [5]. Ishii and Ullmer, who suggested and established the term TUIs, define them as "devices that give physical form to digital information, employing physical artifacts as representations and controls of the computational data" [22]. Both of these definitions highlight the mapping between the physical object and the digital information or function it embodies as the essence of a TUI.

Successful TUIs will therefore contain successful physical/digital mappings. Since we are spatial beings and TUIs are spatial objects, successful mappings will exploit knowledge of how humans act in their spatial environment. We find this knowledge in previous research on human interaction, perception and activity theory. We isolate from this knowledge three heuristics. First, physical/digital mappings must be successful spatial mappings. That is, the relationship between the spatial characteristics of the TUI's objects and their use must be spatially congruent and/or well-known. Second, physical/digital mappings must unify input and output space. Here, we advocate that the purely digital distinction between both input and output spaces and input and output devices must be eliminated. Finally, physical/digital mappings must enable trial-and-error activity. To allow unconstrained exploration of digital problems in physical space, successful mappings will consist largely of fixed one-to-one mappings, rather than transient many-to-many or one-to-many mappings. In what follows we will discuss these heuristics in more detail, and using them, review several existing TUIs. In so doing, we find that the most successful TUIs are "spatial TUIs", a subset of TUIs that mediate interaction with shape, space and structure.

## 2. Spatial Mapping

 "…To primitive man each thing says what it is and what he ought to do with it: a fruit says, 'Eat me'; water says, 'Drink me'; thunder says, 'Fear me'…"

Kurt Koffka, *Principles of Gestalt Psychology*, 1935 [12]

Humans can use and manipulate most objects in the physical world in a natural and relatively effortless manner. We deduce the inherent functionality of objects from their physical qualities – qualities such as shape, weight, size and color. An object's functions as expressed through its physical form are called its "affordances" [6,8,13]. Along similar lines, Norman also discusses a "natural mapping" between an object and its functionality. A clear natural mapping means that the object's functionality is obvious from its physical and spatial characteristics, based on "physical analogies and cultural standards" [13]. Natural mapping is a "primitive" quality of objects and exploits our schooling as spatial beings. Once we are familiar with the spatial qualities of an object we can often map these qualities to new functions and tasks, effectively creating new intuitive spatial mappings.

Beaudouin-Lafon's work helps quantify the spatial congruence between a physical object and its use in a digital application. His "degree of integration" is the ratio between the degrees of freedom (DOF) of an object and the DOF of its application [1]. For example, the mouse affords two degrees of freedom or dimensions of motion, while the cursor does as well – a good ratio of one. Beaudouin-Lafon also discusses a "degree of compatibility" that measures the similarity between each of the actions performed on an object, and its application response [1]. With a mouse, moving the mouse to the left moves the cursor to the left. The mouse is not as well-suited to controlling image plane rotation, though it can be modified to afford this action (see [24]). Objects in the physical world usually have ideal degrees of integration and compatibility, or at least are perceived to have these qualities by the human who uses them. This is because in the physical world, objects and their function are usually unified and inseparable, making the effect of their manipulation intuitive and easy to anticipate.

Spatially incongruent mappings can also be quite efficient, if they are well-learned. A classic example is the QWERTY keyboard, which maps the largely non-spatial alphabet to a spatial keyboard layout. In interfaces, designers can exploit these previously learned mappings. We believe that the quality of such interfaces will depend strongly on how well the incongruent mapping has been learned, with mappings learned early and practiced often being the best candidates for use in interface design.

Collectively, these qualities define what we call the spatial mapping, or the spatial relationship between a physical object and its digital use. Everyday objects typically offer a clear and intuitive spatial mapping to their function – sometimes so clear that people forget the mapping exists. In human-computer interfaces the mappings are often much more complex and profoundly limited by the affordances of the physical interface components, resulting in an unintuitive and frustrating interaction experience for the user. One example is the mapping of some high DOF gaming applications to the two-dimensional mouse, using mouse buttons to select manipulation of different DOFs. Most users find such mappings frustrating, and switch to a keyboard interface. TUIs, with their heavy reliance on physical objects, are even more spatial than most interfaces. Therefore a good spatial mapping is crucial if they are to succeed. In our opinion, good spatial mappings will be achieved most easily if the application itself is inherently spatial – that is, if it mediates interaction with shape, space or structure. When coupled with TUIs, we call these sorts of applications "spatial TUIs". We will review several good examples of spatial TUIs below. Of course, TUIs can be mapped to digital functions that are not spatial, such as Boolean queries of databases. We argue that although such mappings are possible, the resulting interfaces will have many hurdles to overcome if they are to succeed.

Of course, choosing a spatial application is only a good beginning for TUI design. Spatial congruence should be maintained in the mapping by ensuring good degrees of integration and compatibility, matching DOF number and type in the physical interface and digital application. Failing this, learned mappings should be exploited – the more primitive or "hardwired", the better. And even when this is done, affordances should not be forgotten. Both TUIs with good degrees of integration and compatibility and those built on well-learned non-spatial mappings can still have poor mappings, simply because they do not afford their use. For example, if we miniaturize an otherwise useful TUI by a ratio of 1 to 10, we will find that although it might excellent by all our guidelines, it is too small to afford any human interaction.

## 3. I/O Unification

In the natural, physical world we make little distinction between input and output, with perhaps the closest approximation being cause and effect. Only when working digitally is this distinction fully introduced. We see two highly harmful consequences to this split: decoupling of action and perception space, and uncertainty about state. TUIs are capable of overcoming these difficulties by unifying input and output. When we work with physical objects, we perceive both our fingers and the objects they handle in the same time and space. This coupling of action space (our hands) and perception space (the view and weight of the object) allows us to direct our attention to one time and place [6,7]. In human-computer interfaces, action and perception space are usually decoupled. For example, action with the mouse happens at one location, perception from the display screen at another. This makes our work more difficult by dividing our attention and forcing us to map one space to the other. TUIs allow close coupling of action and perception space, improving user interaction. As physical objects, TUIs naturally provide a tactile fusion of perception and action. Many TUIs also display output on the input surface, unifying input and output visually, and strengthening action-perception coupling further. TUI components are also often mapped to individual digital objects (unlike the mouse, which is continuously coupled to and decoupled from various digital objects), bringing perception and action space into still closer agreement.

The state of activity in the physical world is usually embodied in the tools and workplace being used. When working, humans exploit visual, tactile and other sensory cues to deduce the state of their activity and its progress from the condition and motion of their tools. For much the same reasons, the need to maintain clarity of application state is well established in HCI [20]. Users need good knowledge of state to monitor the progress of their work. Classic HCI solutions to this challenge are visual, relying on the display to output changes to application state according to the input device state. Though this application feedback is crucial, the separation between input and output devices introduces some uncertainty about state: which feedback is more trustworthy, that from the input or that from the output device? By long convention, users have learned to trust display feedback most.

We believe TUIs should push clarity of state still further. As TUIs improve action-perception coupling, they increase user identification between physical interface components and digital application objects. Users expect the physical input components to mirror the state of the corresponding digital objects completely, effectively eliminating the distinction between input and output devices. Achieving this I/O unification makes application state even more clear.

To sum up, we argue that TUIs designed to maximize input and output unification will be more effective because they more effectively exploit human spatiality. To achieve this unification, future TUIs should couple action and perception space and embody a clear representation of state across all sensory modalities. This is best achieved by eliminating the distinction between input and output devices. Ideal I/O unification can require a feedback loop that is able to physically deform and actuate TUI components, without rendering them too cumbersome or hindering their essential utility as input devices. Technologically, this is often easier said than done but it is certainly a worthy goal.

## 4. Support of "Trial-And-Error" Activity

When using physical tools, humans perform activity that is cognitive and goal related as well as and physical and exploratory. Such activity can be deconstructed and studied using the concepts of pragmatic and epistemic actions [5-7,10]. Pragmatic actions are the straightforward maneuvers we perform to reach our cognitive goal. On the other hand, we carry out epistemic actions using the physical task space itself in order to improve our cognitive understanding of the task. For clarity's sake, we speak of "trial-and-error" rather than epistemic actions. Trial-and-error actions often fail to bring us any closer to our goal, but can sometimes reveal completely unexpected information and shortcuts that would have been very difficult to find by following a straightforward, pragmatic approach [10].

A good physical tool enables users to perform pragmatic, goal-oriented activity as well as trial-and-error activity, and ensures that the cost of speculative exploration of the task space is low [6]. This is typically achieved by using tools and workspaces composed of many independent components. This abundance of components makes them detailed and easily manipulated representations of the cognitive task's state, and provides excellent support for trial-and-error activity.

Traditional human-computer interfaces fall short of this standard, since they are designed primarily to support pragmatic actions. These interfaces usually have no persistent coupling between physical and digital objects, forcing users to couple and decouple them sequentially. Trial-and-error activity is then best supported through manipulation of the sequence of couplings and decouplings, using operations such as "undo". These solutions are often inflexible, raising the cost of trial-and-error exploration. For example to undo a single erroneous action, users typically also have to undo all the actions that followed it – even if they were viewed by the user as beneficial. Such approaches can be adequate for text editing but can limit interaction

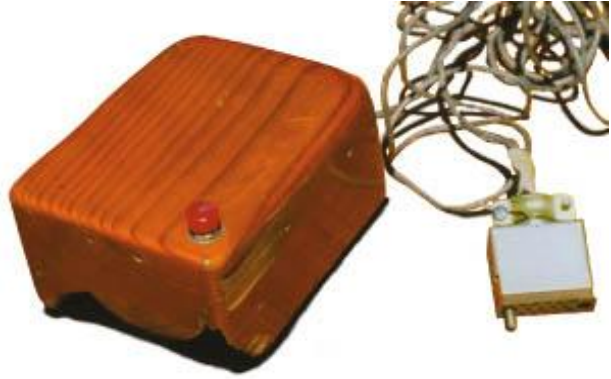

Fig. 1. The first mouse [2]

severely in more complex tasks such as three-dimensional modeling.

TUIs are far more physical than traditional human-computer interfaces, and so can offer much better support for trial-and-error activity. In forming their spatial mapping, TUI designers should strive to provide as many one-to-one couplings as possible between digital application and physical interface objects. Doing so enables a clear representation of the task state, and provides multiple points of access for manipulating and exploring it.

## 5. Heuristics at work

Is this physical, tool-based approach to TUIs useful? With usability studies of TUIs just beginning, a definitive answer to these questions is not available. For now we demonstrate and test our heuristics by analyzing several successful spatial TUIs, moving generally from least to most spatial and starting with the mouse.

### 5.1 The Mouse

Is the mouse (Figure 1) a TUI? Certainly the mouse's undisputed success should be attributed to its engagement of human spatiality. According to Ishii and Ullmer's definition, when considered only as a means of manipulating the cursor in a desktop interface, the mouse is a TUI. Here, the mouse is a physical object used to control a digital object, giving physical form to digital information. However, as it is more generally used, the mouse is not a TUI by that definition. At times the mouse becomes a paintbrush, a view direction and a page-turner. Clearly this single physical object cannot reliably suggest the physical form of all these digital objects. Let us examine the mouse further in light of the TUI design heuristics we have presented.

The quality of the mouse's spatial mapping depends on its application. The mouse is highly spatial; it is physically easy to hold it and roll it on a surface. At times, the mouse's mapping is very intuitive, for example when moving a cursor, selecting and moving windows, and in numerous other planar tasks. However, the mouse is used for many higher dimensional tasks, from changing the view in a first person shooter to editing three-dimensional structures with CAD software. In such applications, the mouse has poor degrees of integration and compatibility, and cannot claim to rely on any previously well-learned non-spatial mapping.

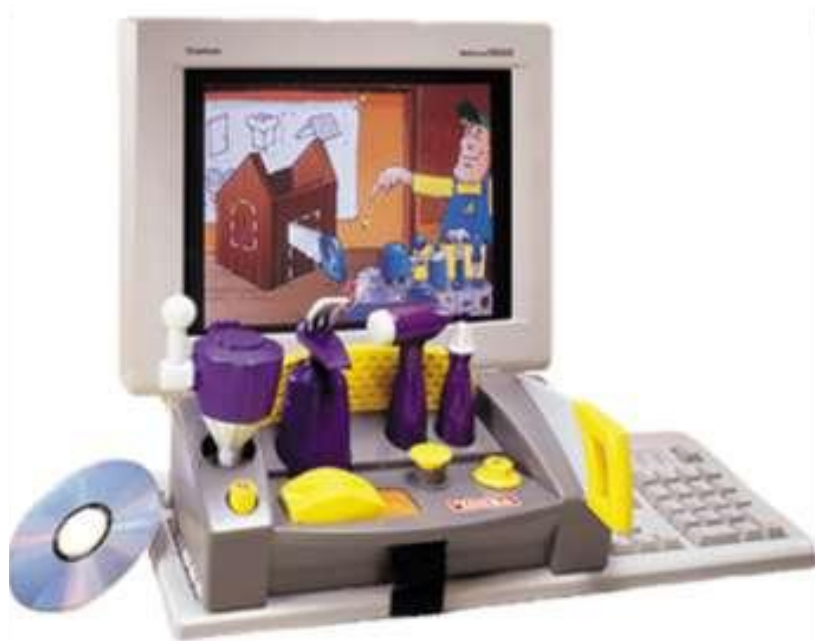

Fig. 2. Tonka Workshop [21]

The mouse provides neither good I/O unification nor trial-and-error support. Application state is not reflected in mouse state, since display occurs in a separate device. The mouse separates action space (the hand moving the mouse) from perception space (the cursor moving on the display). As a single physical pointing device for many digital objects the mouse also cannot offer one-to-one coupling, raising the cost of physical exploration of a problem space.

To conclude, the mouse does not offer strong spatial mapping to many of its applications. It also makes a very poor TUI by the rest of our heuristics. Given these limitations, the mouse's great success suggests the great potential for more ideal TUIs.

## 5.2 Tonka Workshop

Tonka Workshop (Figure 2) is a simple and low-cost toy [21]. It can also be seen as an example of a mature and commercially successful spatial TUI.

Tonka Workshop consists of a set of physical workshop tools, including a saw, mallet, screwdriver, hand drill, and spray can. The TUI is attached to the top of a regular keyboard and translates the actions performed with the tools to keystrokes. The user actions that can be performed with the tools are restricted by the types of tools available, and the overall simplicity of the Workshop interface. For example some tools cannot be picked up, and even when they can be, tool action is limited to pre-defined locations on the Workshop input surface. However simple, the TUI creates a convincing interaction experience for its target users. The tools offer simple, spring-based passive haptic feedback when actions are performed with them, and Workshop reinforces this feedback with additional audio and visual feedback. For example using the saw will generate a sawing sound and some dust (which the user can later clean with a broom). The software challenges young users with a variety of tasks, including household chores, constructing spacecraft, fixing a PC and sculpturing metal and stone.

The Tonka Workshop does not support I/O unification, with display and interface strictly separated and the physical TUI barely reflecting task state. The TUI's multiple physical tools with their clear one-to-one mappings to the virtual task can potentially support trial-and-error activity. However, trial-and-error is hardly supported in the current Tonka Workshop implementation. Nevertheless, Tonka Workshop's spatial mapping is extremely intuitive and simple. The resulting TUI enables very young users (including two-year-olds) to achieve levels of interaction that would be impossible with standard interfaces like the mouse.

## 5.3 Monkeys

Monkeys (Figure 3) are a class of devices for inputting articulated poses, used in the film industry [3]. These devices enable animators to manipulate three-dimensional digital models into desired keyframe postures. Each of a Monkey's joints contains a sensor that measures joint angle. Because the Monkeys' skeletal topology and geometry are well known, the complete pose of a Monkey can be determined using only these angles.

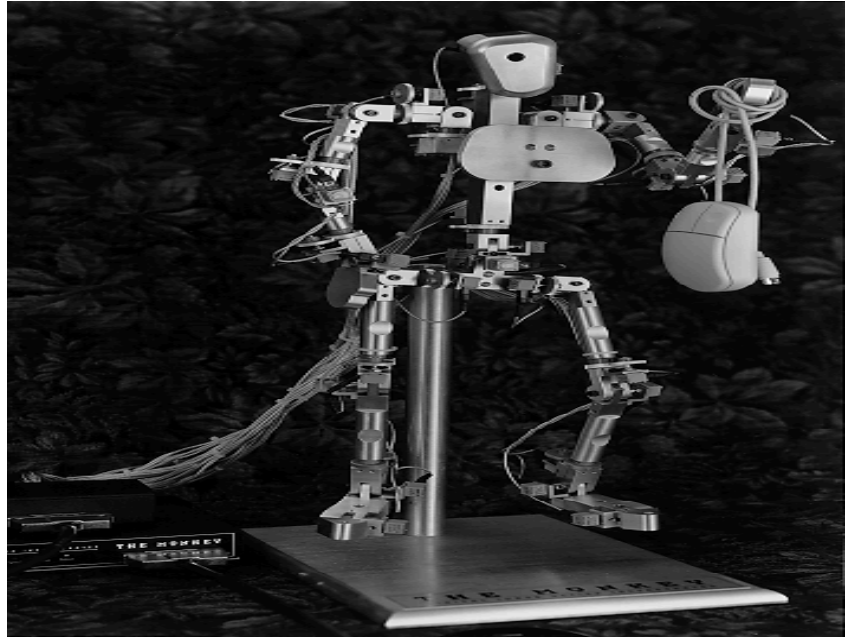

Fig. 3. Monkey (holding a mouse) [3]

Monkeys are very effective spatial TUIs, introducing an intuitive interface for a specialized task that can be extremely difficult to perform with traditional mouse-based interfaces. Monkeys use very congruent spatial mappings, with each digital DOF mapped to a single physical DOF and each such component mapping highly compatible. Strictly speaking, Monkeys do not unify input and output, with the virtual pose being presented on a screen separate from the physical device. However, we can argue that Monkeys themselves serve as physical displays, reflecting task state by their mere existence and unifying input and output (of course, any abstraction or sampling error introduced in the physical/virtual mapping would weaken this argument). Monkeys fully support trial-and-error activity, representing the digital problem space completely in the physical interface and allowing the animator to access the problem at any point, in any order.

Yet a Monkey's strength is at the same time its weakness. The highly specific spatial mapping offered by Monkeys usually means that each digital character requires its own Monkey, limiting the generality of the interface. Using an old Monkey with a new digital character having different limb lengths can be extremely frustrating, even if the DOFs are unchanged – the old Monkey simply does not offer the same affordances.

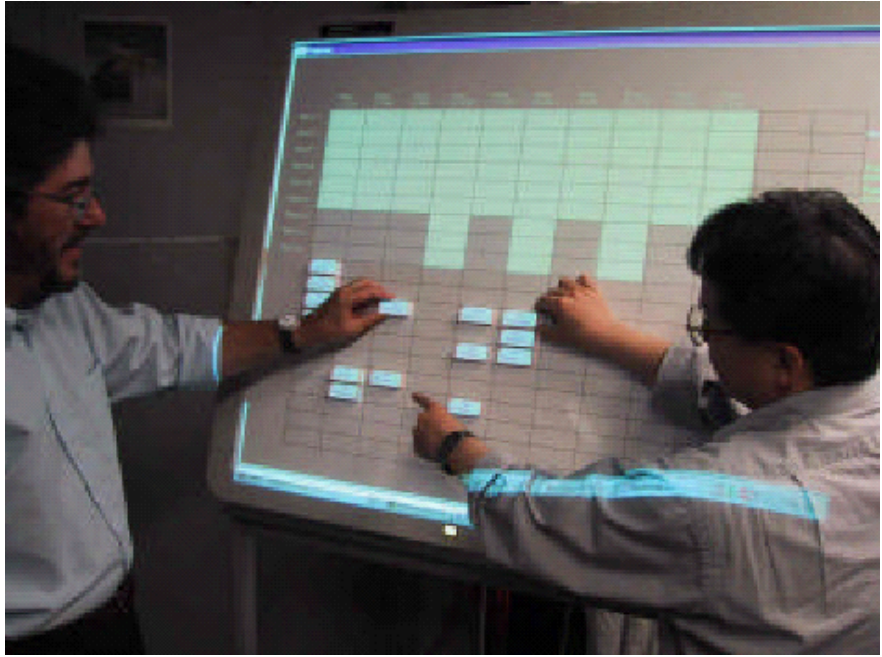

Fig. 4. Senseboard [9]

## 5.4 Senseboard

Jacob et al.'s Senseboard (Figure 4) was designed for information organization [9], such as scheduling. These tasks are commonly performed using spatial placement and manipulation of physical tokens or representations, for example in scheduling, moving paper tokens representing events on a timetable. Senseboard was designed to support this physical approach to information organization and enhance it with automation.

Senseboard consists of a vertical surface augmented with digital projection. On the surface, the user can place rectangular, magnetized pucks containing RF tags that enable Senseboard to track their ID and location as well as augment their appearance.

Some of the pucks serve as "Data Objects", representing a component of the information being organized, while others are "Commands" representing an action that can be executed on Data Objects.

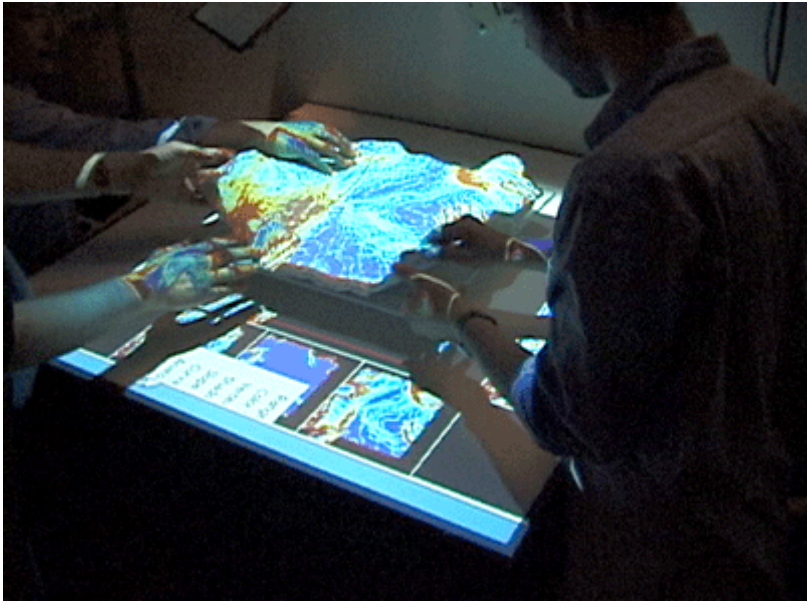

Fig. 5. Illuminating Clay [16]

Through automation the TUI enables a set of actions that were not feasible in the paper-based approach. The timetable is constantly tested for conflicts, and if these occur they are highlighted. Entries can be easily grouped or ungrouped, present a summary or detailed view of the information they represent, and the entire draft schedule can be easily saved.

Is Senseboard's spatial mapping intuitive? We believe that while the mapping cannot be called congruent (information is not naturally spatial), Senseboard brilliantly exploits an incongruent spatial mapping that is learned relatively late in life. Users accustomed to scheduling tasks, timetables and the use of physical tokens for these tasks will find Senseboard to be a natural interface.

Senseboard provides excellent support for trial-and-error activity. In fact even before automation, information organization using tokens is already highly epistemic. With both paper tokens and Senseboard, any schedule may be tested by simply moving tokens or pucks on the timetable. Should that schedule prove unsatisfactory, new schedules may be tested by reversing or changing any previous action. Senseboard unifies input and output by projecting information on top of the interaction surface and the pucks. The visual illusion of the projected information integrated with the physical tokens is effective, but is still a long way from a strong physical binding. The visual illusion can be broken by shadows from the user hands, latency (delay in system response), disappearance of the projected details when the puck is taken off the TUI and the inability of Senseboard to move pucks when it has an alternative position to suggest.

## 5.5 Illuminating Clay

Piper et al.'s Illuminating Clay (Figure 5) is a landscape analysis tool based on interaction with a physical clay model [16]. During interaction Illuminating Clay visually augments the clay model with various real-time landscape functions, presenting information such as slopes, shadows, solar radiation, land erosion and water flow.

Much like Senseboard, Illuminating Clay augments the interaction surface by projection. However quite uniquely, Illuminating Clay enables interaction with a three-dimensional surface, using a laser scanner to capture the surface topography of the clay model in real time. The surface of the clay model itself is the interface – it can be spatially manipulated to achieve the application goals, even without placing of physical objects on it.

Illuminating Clay offers an intuitive spatial mapping to its task. Although the clay model differs from the landscape it represents in scale and in the tools required to alter it (hands versus heavy machinery), the mapping is clear and well practiced from an early age (in this respect, see a variant of Illuminating Clay, SandScape, in which sand was used instead of clay [23]).

Illuminating Clay unifies input and output in much the same way as Senseboard does, with the same strengths and limitations. We can argue however that the three-dimensional spatial clay model surface offers a consistent representation of a major feature of the task state and hence a stronger physical fusion of input and output. Trial-and-error actions are naturally supported since the clay captures the complex state of the landscape well, holds its shape and thus maintains state well, and offers myriad points of access to this state on the clay, making it simple for users to explore the task space physically.

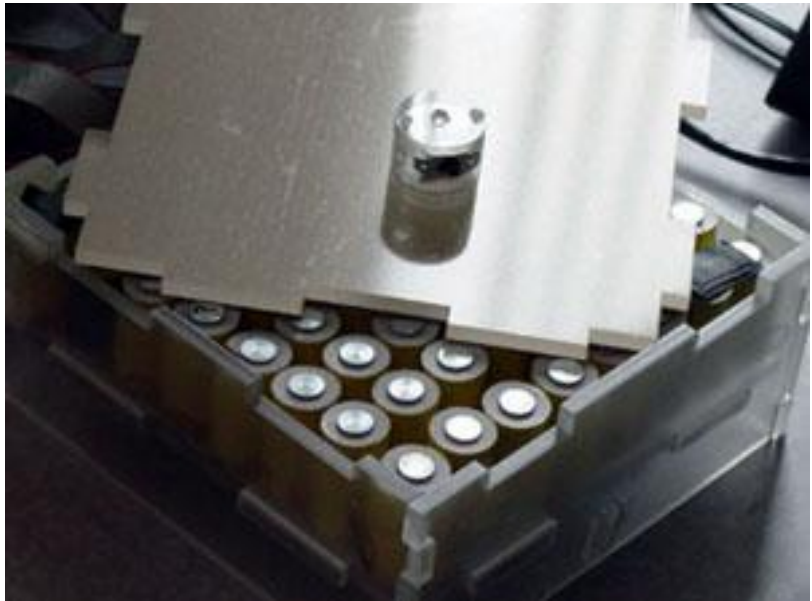
Fig. 6. The Actuated Workbench [15]

## 5.6 Actuated Workbench

The Actuated Workbench (Figure 6) is an interesting attempt to unify I/O not only to the visual sense, but also to the tactile sense [15]. Currently more of a conceptual interface prototype than an application interface, Actuated Workbench mediates interaction on a surface using user-controlled pucks, while the TUI augments the surface with visual display. Most interestingly, Actuated Workbench can move the pucks itself to reflect internal changes to task state. Under the Workbench surface is an 8X8 grid of electromagnets that can manipulate puck position (the designers also mention the possibility of flipping pucks).

Since Actuated Workbench has not been coupled with a task we cannot evaluate its spatial mapping, only potential mappings – and with the abundance of tabletop TUI applications that potential seems great. As with most tabletop TUIs, the Actuated Workbench's support of multiple points of access enables trial-and-error actions when exploring its task space. Most uniquely, the Actuated Workbench unifies input and output in the tactile realm. The possibility of outputting internal state using the physical manipulation of interaction mediators is intriguing and still largely unexplored.

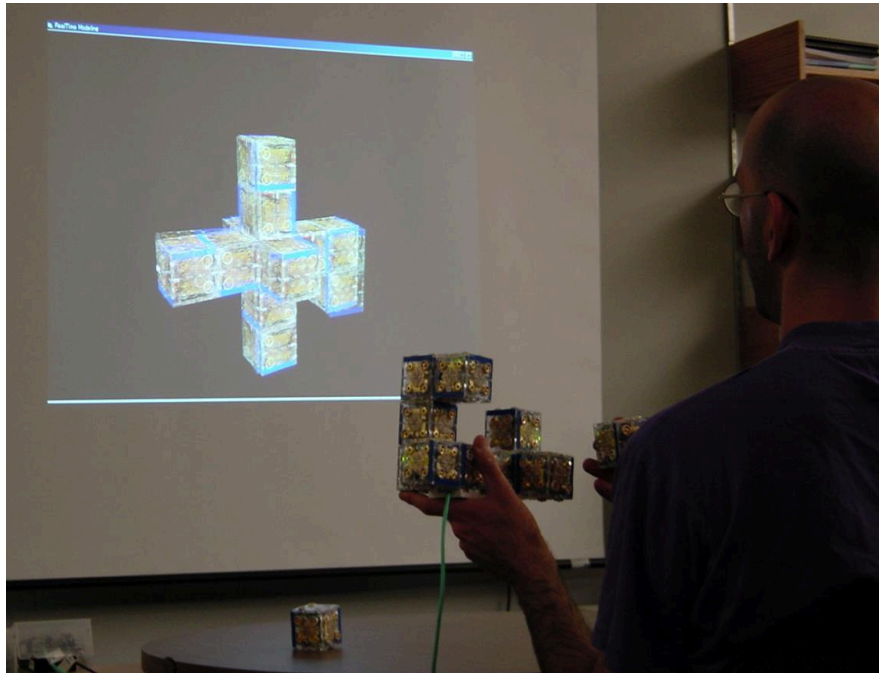

Fig. 7. Cognitive Cubes [19]

## 5.7 Cognitive Cubes

Cognitive Cubes (Figure 7) is a system for cognitive assessment of human constructional ability [19]. Cognitive Cubes follows a very simple assessment paradigm: show participants a prototype, and ask them to reconstruct it. In Cognitive Cubes, the prototype is an abstract three-dimensional shape constructed of simple blocks and displayed visually, while participants attempt the reconstruction using physical versions of the same blocks. During this attempt, each change of shape is automatically recorded and scored for assessment.

Cognitive Cubes is based on ActiveCube [11], a spatial TUI that supports three-dimensional construction using a set of plastic cubes (5 cm/edge). The cubes can be attached to and detached from other cubes on any of their six faces. Connections form not only a physical shape but also an electrical network topology. A host computer regularly samples the network and the resulting shape, registering connection and disconnection events in real time.

Cognitive Cubes offers a very intuitive spatial mapping between TUI and task. Constructional assessment activity is performed entirely in the physical domain using ActiveCube, which naturally affords constructional activity, much like Lego blocks. Degrees of integration and compatibility are ideal.

At first glance Cognitive Cubes does not seem to offer strong I/O unification, because the virtual prototype is displayed separately from the participant's current physical approximation. However, the prototype is merely an unchanging representation of the constructional goal, and is therefore external to the interaction. A tighter coupling between the prototype and approximation would make the assessment task surprisingly trivial! However, Cognitive Cubes' representation of the user's approximation is completely unified – so unified in fact that it is not necessary to "output" a representation of that approximation.

Lastly, like many other construction sets, Cognitive Cubes offers excellent support for trial-and-error exploration of the problem domain. Participants can perform actions on any part of the problem state in any desired order, undoing their former actions in a very flexible fashion.

## 6. Conclusion

Like any sort of human-computer interface, TUIs strive to improve human productivity by making the full power of automation accessible to users. We maintain that TUIs achieve this accessibility by tailoring the physical form of the interface to users' tasks, exploiting humans' innate spatial skills. From this perspective, the coupling between the physical and digital components of a TUI is of secondary importance. The fundamental quality of a TUI is determined by the coupling between the TUI and the task it is designed to support.

One potential shortcoming of our approach is the resulting specialization. Tailoring of the physical interface to tasks to the extent we advocate implies poor tool generality, which contrasts dramatically with the extreme generality of current computers and their interfaces. More generic TUIs (see for example the GUI/TUI hybrid in DataTiles [17]) might be able to increase accessibility to automation across a wider variety of tasks.

Nevertheless, we believe that highly specialized TUIs will prove to be more valuable than generalized TUIs in the long run. Many before us have pointed out that today's PCs are complex devices that are difficult to master (for example, see [14]). The unspecialized PC interface leaves many applications and user groups poorly served. Addressing this problem will require replacing today's generic PCs with specialized information appliances, creating a growing need for a variety of customized TUIs that will complement or replace the standard generic interfaces.

Successfully specialized TUIs should strive to follow these guidelines:

1. Construct an intuitive spatial mapping to the application task. Exploit spatial abilities and mappings known innately and learned early in life before those learned later.
2. Further exploit users' spatial abilities by supporting trial-and-error actions, and unifying input and output. This maximizes the usefulness of the mapping by reproducing the real-world spatial setting.
3. Explore the rich real-world vocabulary of physical objects, tools, and related spatial techniques as inspiration for novel TUI design.

In the near term, researchers should increase the number of spatial TUI applications. In the process, we will benefit each applied field and learn a great deal about successful TUI design. For example, to enable new prototyping and spatial assessment applications, computerized construction sets similar to ActiveCube but with the size of regular Lego blocks would be extremely useful.

The principal longer term challenge is the unification of input and output. As we discussed earlier visually augmenting the input space can be effective, but is limited by the failure to unify input and output in the non-visual sensory modalities. We believe that a combination of visual augmentation with automated physical control of the TUI's input space could become a powerful technique used in successful spatial TUIs. One important example of such a unifying technology would be "digital clay" (see for example [18]). Digital clay would allow users to input detailed three-dimensional shapes using sculpturing techniques, and at the same time support physical output that interactively changed shape according to application need or state.

Given the technical difficulty of implementing I/O unification, it is interesting to note the relative ease with which trial-and-error actions can be supported. Offering multiple, stable and easily manipulated points of interface access to the task state is much simpler than finding or designing a technology that can act simultaneously as a visual display, visual input, haptic display and haptic input device.

The tangible user interface is still a very young technology and field of research. We look forward to its further development and application with excitement.